# A CAPABILITY APPROACH TO AI ETHICS

Emanuele Ratti[a]

Mark Graves[b]

**Abstract**. We propose a conceptualization and implementation of AI ethics via the capability approach. We aim to show that conceptualizing AI ethics through the capability approach has two main advantages for AI ethics as a discipline. First, it helps clarify the ethical dimension of AI tools. Second, it provides guidance to implementing ethical considerations within the design of AI tools. We illustrate these advantages in the context of AI tools in medicine, by showing how ethics-based auditing of AI tools in medicine can greatly benefit from our capability-based approach.

## 1. INTRODUCTION

In the past few years, AI ethics has emerged as an important scholarly field to shape the governance of AI. The number of articles claiming to contribute to AI ethics is impressive (Tsamados et al 2021). Because of this sudden growth, AI ethics has not had yet the opportunity to stabilize itself as a discipline with its own identity, and there have been attempts to either import already existing frameworks from other practical ethics discipline (most notably, biomedical ethics), or to concentrate on more technical solutions, while leaving ethical reflection at a minimum, as in the case of initiatives of fair-ML (Fazelpour and Danks 2021). In this article, we attempt to build a foundation to AI ethics by adapting the capability approach (Section 2) to the specificities of the AI context (Section 3). In particular, we show that conceptualizing AI ethics through the capability approach can solve challenges related to the 'applicability' of AI ethics in actual practice (section 3.2), as well as addressing more theoretical issues affecting current AI ethics approaches (section 3.1). We apply the considerations of this article to a case of ethics-based auditing of medical AI tools (Section 4).

### 1.1 The context

Before elaborating our integrative framework, we need to clarify a few aspects of the context of this article. First, given that the term AI covers different kinds of techniques (Cowls and Floridi 2019) and the impressive variety of contexts in which AI is used, we necessarily need to restrict our scope. As for the meaning of AI, we refer to a particular subset of AI tools, namely data-centric AI and machine learning drawing upon data science processes (Ratti and Graves 2021). As for the context of use, we will refer to the medical context, especially in Section 4.

---

[a] Department of Philosophy, University of Bristol, correspondence author: mnl.ratti@gmail.com
[b] AI & Faith, San Francisco

Second, it is important to address from the start the role of ethics in the regulation and governance of any kind of technology, and AI in particular. We do this through the distinction between hard and soft governance and the role of ethics in them (Floridi 2018). On the one hand, *hard governance* refers to systems of rules or regulations which are enforced through public institutions, from governments to city councils. In the context of AI, hard governance has been prioritized, and for very good reasons, given the increasing power of corporations like Google or Amazon and the lack of regulatory oversight on systems that have profound impacts on citizens' lives (Zuboff 2015). On the other hand, *soft governance* refers to either behaviors that allow some degree of flexibility, or to behaviors not covered by hard regulations that nonetheless have important societal ripple effects, or to social norms existing outside legal frameworks. Soft governance of AI is a timely issue, and it covers a wide range of topics, including issues on the nature of explainability of AI, trust, online manipulation, accountability, etc.

AI ethics can play two roles, depending on whether one considers it as contributing to either soft or hard governance. First, it can be used as a foundation for hard regulation; after all, regulations are also shaped by considerations concerning values, duties, and rights (among the many things). Let us call this *hard ethics*. But ethics can be also fruitful for navigating the territory not covered by hard regulations. As Floridi puts it, ethics can be used as a tool to understand "what ought and ought not to be done over and above the existing regulation" (Floridi 2018, p 4) – in other words, ethics can be used as a 'post-compliance' soft governance tool. Let us call this *soft ethics*. This article will deal with *soft ethics*, where the goal is to align AI practitioners' 'day-to-day' activities to ethical considerations, over and above the aspects of compliance to the law.

There have been various attempts to provide a solid foundation to AI ethics, and these have noteworthy consequences on how fruitful hard and soft ethics have been. Some have thought to import the main tenets of principlism (Beauchamp and Childress 2009) from biomedical ethics.[1] This has been especially successful for ethics contributing to hard regulations, but it is more problematic in the context of soft governance, in particular for the difficulties of translating principles into actionable insights that can be concretely used by AI practitioners in their day-to-day activities (Morley et al 2019; Bezuidenhout and Ratti 2021). A different trajectory has been proposed for soft ethics, one that is focused on the efforts of researchers working on fairness, accountability, and transparency for machine learning tools (Fazelpour and Danks 2021). However, these efforts tend to provide off-the-shelf tools that require practitioners to make value-laden choices among different measures, but making these choices requires "philosophical arguments and considerations that fall outside of the narrow technical scope of the standard approach to fair ML" (Fazelpour and Danks 2021, p 10). This means that one dimension of soft ethics - namely, ethical reflection - is not addressed by the fair-ML perspective. In this article, we propose to conceptualize the foundations of AI ethics, and soft ethics in particular, through the lens of the capability approach (Sen 1999; Nussbaum 2011). As we will show, the capability approach is successful where principlism and fair-ML have failed (i.e. in translating ethics into actionable insights, and in fostering ethical reflection).

## 2. THE CAPABILITY APPROACH AND TECHNOLOGY

Because of the centrality of the capability approach in our framework, in this section we introduce its main ideas, and we emphasize relevant aspects of this approach, especially the ones connected to technology, given that AI is a kind of technology.

### 2.1 The capability approach

The capability approach is a strategy to provide comparative quality-of-life assessments for "theorizing about basic social justice" (Nussbaum 2011, p 18). In particular, it is used to evaluate the impact of social arrangements and policies on individuals; and it does this in a flexible way and in a variety of contexts, from policies of affluent societies, to the ones designed for low- and middle-income countries (Robeyns 2005). The capability approach has been theorized in various ways, starting from Sen's initial formulation (1985). Here we will especially consider Nussbaum's formulation (2006; 2011).

The cornerstone of the capability approach is that individuals not only should have access to the concrete positive resources necessary for improving their quality of life, but also "they should be able to make choices that matter to them" (Alkire 2005, p 117). This idea has one important ramification (among the many), which is the fact that we should evaluate policies or institutions not only on the basis of what individuals concretely get as a result of the application of that policy; also, we should evaluate what people can possibly do or be as a result of the policy itself. In other words, the focus of policy evaluation should be especially on those aspects that either obstruct or facilitate the fulfillment of "the kind of life that [people] have reason to live" (Robeyns 2005, p 94). From these basic considerations, an important distinction is made, which is the one between functionings and capabilities. As far as we know, all supporters of the capability approach accept this distinction.[2]

Functionings are achievements, or "the various things a person may value doing or being" (Sen 1999, p 75). These vary substantially, from something basic like being nourished, to something quite sophisticated such as participating in a strike. The capability approach differs from other (and more consequentialist) approaches to welfare in that it forcefully argues that measuring life's quality only on the basis of functionings is misleading. In order to have an accurate representation of one's life quality, we need also to consider the freedom that an individual has in deciding what functionings to achieve and how. A typical example used in the capability literature is the one about the difference between fasting and starving (Alkire 2005). From the point of view of functionings, those who fast and those who starve are alike: they both lack nutritional intake. But the freedom that fasting individuals have in deciding whether to fast and how indicates a rather different quality of life than the one experienced by those who starve. The right question to ask is not just what a person does or has but rather what a person is able to do and be.

These things that a person 'is able to do and be' are called *capabilities*. They are defined as a range of potential functionings that are concretely feasible for a person to achieve, and that one can freely choose to achieve - "a set of (usually interrelated) opportunities to choose and to act" (Nussbaum 2011, p 20). Capabilities are also said to be *substantial freedoms*, given the importance and the focus on choice and opportunities "which people then may or may not exercise in action: the choice is theirs" (Nussbaum 2011, p 18). These substantial freedoms have a remarkable ethical dimension, as they are conceived to provide a foundation for human dignity, at least in Nussbaum's formulation. It is also worth noting that the Capability Approach is not a theory that 'reads norms off' of a concept of human nature. As Nussbaum (2011) argues, the capability approach "is evaluative and ethical from the start" (p 28). This evaluative dimension raises the question of which are the most important capabilities. The issue is important because we may want to foster human flourishing by expanding capabilities, but not all capabilities' expansion will result in more flourishing. For instance, the capability to become a hitman is not just irrelevant to human flourishing; in fact, it is not a difficult endeavor to show that the capability of becoming a hitman is in tension with human flourishing in the first place. Nussbaum invokes explicitly the notion of human dignity and argues that her version of the capability approach "focuses on the protection of areas of freedom so central that their removal makes a life not worthy of human dignity" (2011, p 31). Starting from what 'bare minimum' we should guarantee for human dignity, Nussbaum formulates a list of ten 'central capabilities' (2011, pp 33-34). The task of a government, in her view, is to make sure that all citizens are secured a threshold level of these (even though it is not clear what this threshold is). Interestingly, these capabilities are also referred to by Nussbaum as political liberties, and when a society overrides these liberties, it "has delivered to its members a merely animal level of satisfaction" (1997, p 291). Surely the list can be expanded, but her approach - in our interpretation - requires that no central capability can be removed from that preliminary list without violating human dignity.

## 2.2 Capabilities, conversion factors, and technology

Whether one is in the position to pursue a specific functioning in the way they value, depends on what are called *conversion factors*; that is, those factors that facilitate for a person to turn an abstract possibility into something actual (Robeyns 2005).

There are at least three types of conversion factors (Robeyns 2005). First, there are *personal conversion factors*. These are characteristics of individuals that are necessary to pursue certain functionings, such as a proper metabolism, reading skills, physical characteristics, etc. In our interpretation, these personal conversion factors include, in Nussbaum's terms, both 'internal capabilities' ("trained or developed traits and abilities" 2011, p 21) and 'basic capabilities' ("innate powers that are either nurtured or not nurtured" p 23). A second category are *social conversion factors*, namely those social conditions facilitating/restricting the concrete exercise of substantial freedoms; these include things like social norms covering discriminating practices, gender roles, formal and/or informal hierarchies. Finally, there are *environmental conversion factors*, which include climate and geographical location. It is worth emphasizing that Nussbaum calls the combination of all these factors 'combined capabilities', namely "the freedoms or opportunities created by a combination of personal abilities and the political, social, and economic environment" (2011, p 20).[3] In our opinion, the idea of combined capabilities reflects the fact that capabilities cannot really be separated from the conditions that make them concrete possibilities of pursuing functionings. Or better: they can be separated analytically (as we are doing now), but in designing policies with certain (capabilities-based) aims, we cannot consider the capabilities separated from the conversion factors that make them concrete possibilities.

In order to understand the connection between conversion factors and capabilities, consider a fictional (but realistic) example. Imagine a policy which makes mandatory for young boys and girls to go to school until they are eighteen. The policy's goal can be formulated in terms of the expansion of capabilities of, e.g., senses, imagination, and thought, practical reason, affiliation, etc. But whether the policy is successful in actually expanding those capabilities for individuals will depend on conversion factors. For instance, there might be social norms in certain neighborhoods that do not see in a particularly favorable light young girls going to school until they are eighteen. Or in certain areas there might be geographical obstacles to school access in the first place. Because of the presence of certain social and environmental factors (resulting in the lack of the right social and environmental conversion factors), the policy might not expand capabilities in the way policy makers envisioned.

Because AI is a kind of technology, it is important to understand the relation between technology, capabilities, and conversion factors. Intuitively, technology should be important for capabilities, given that artifacts in many cases expand the range of things we can do or perceive. But despite this widely shared intuition, research on the relation between technology and capabilities is not as developed as one might expect. The situation has improved in the last decade, due especially to the work of Ilse Oosterlaken (2013; Oosterlaken and Van den Hoven 2012). It is possible to distinguish at least three ways in which technology can in part shape and/or constitute capabilities. First, technologies are built with the purpose of expanding capabilities. Consider bicycles, which is a typical example made in this context (Oosterlaken 2013). Bicycles can be considered technological artifacts that greatly contribute to the expansion of essential capabilities such as affiliation; sense, imagination and thought; and others. However, the expansion of capabilities operated by bicycles requires the right configuration of conversion factors, exactly like in the above case of school policy. If you live in the Netherlands, then owning a bicycle greatly expands those capabilities, because there is in place a cycling infrastructure that allows you to bike (Oosterlaken 2013). However, if you live in the desert, then that infrastructure might not be present, or the climate might make cycling very difficult, and hence bicycles will not expand capabilities. A second impact of technology on capabilities is more indirect; technology can shape the very "socio-technical systems, institutions, and practices in which we are also embedded" (Oosterlaken 2013, p 88). A classic example is how information and communication technologies (ICTs) have shaped medical institutions in a way that impacts health as a capability. Among the many cases, think about the impact of telemedicine in health - it can certainly reduce costs, but also dramatically change the patient-doctor relationship. Finally, technology can also influence our own interpretation of what it means to expand a certain capability or what a capability really is. For instance, social media platforms have been shaping the way we think about personal relationships with people (Vallor 2016), and hence they are having an impact on conceptions of 'affiliation'.

## 3 CAPABILITIES AND AI

As we have briefly mentioned, any technology can potentially have an effect on capabilities. As a technology, we should expect AI to have an effect on capabilities as well. In this section, we develop an approach to AI – and to AI ethics – that is based on the capability approach[4], and we show that this is fruitful in two senses. First, it clarifies what is the 'ethics' in AI ethics, and hence provides a precise frame for the ethical discourse within a technical context like the one of AI. Second, it is a guide for AI ethics in practice, by providing precise heuristics for identifying the concrete ethical dimension of AI tools, as well suggesting ways in which the ethical dimension of AI tools can be improved in an interdisciplinary way.

## 3.1 The 'ethics' in AI ethics

Existing approaches are often vague in framing what is the 'ethics' of AI ethics, at times taking for granted the ethical dimension of general principles (Jobin et al 2019), or overlapping ethics and law (Bird et al 2020), and sometimes flattening ethics to mere safety (van Wynsberghe and Robbins 2018). The capability approach provides a precise way of talking about 'ethics' in the context of AI that is rich and substantive. Let us see how.

As we have seen, AI tools (as any other technology) can potentially expand or restrict capabilities. To put it differently, AI tools have a profound impact on substantial freedoms in the sense that they structure and constrain the ways humans interact, act within a context, and hence what they can potentially achieve and aspire to achieve. To illustrate this point, consider Danaher's work (2016). In the context of data mining, he refers to 'algocracy', as a system of governance where "algorithms structure and constrain the ways in which humans within those systems interact one another, the relevant data and the broader community affected by those systems" (p 247). The term 'algocracy' comes from Aneesh's work (2006), originally restricted to structuring forms of work performance. By structuring the ways humans interact with each other and the opportunities that individuals have within a context, AI tools shape what individuals can in principle achieve. It is easy to frame the 'mediation' of AI tools described by Danaher and Aneesh in terms of capabilities and substantial freedoms, exactly because AI tools, by mediating our world, function as bottlenecks for our functionings. For instance, intelligent tutoring systems and other AI-assisted applications have drastic effects on education, especially for remote learning. This impacts the capability of senses, imagination, and thought and what functionings (related to that capability) are available for us to achieve. One can (in principle) have access to different educational contents that under normal circumstances might not be at someone's reach because of geographical barriers, but which technology allows. Social media can shape meaningful relations with other people, thereby impacting the capability of affiliation and its related functionings.

How does this help us to better understand what is the 'ethics' in AI ethics? As we have mentioned above, capabilities are substantial freedoms. They are about choosing ways of doing and being that we value and that matter to us. By being about these substantial freedoms and what we value, capabilities shape our own conception of what is the Good Life and the life we feel we ought to live. But ethics is primarily about how we ought to live our own lives, and what the Good Life is, so capabilities are *ethically relevant by definition*. The point is that, because of its scale (Bommasani et al 2021) and ways of slowly becoming invisible (Moor 1985), AI tools shape our own capabilities, and by doing so, shape our substantial freedoms, and hence they are 'ethical' even when they do not do anything 'moral' or 'immoral' – just by operating, they profoundly shape the ethical dimension of our existence. By mediating and shaping our world and environment, AI tools also shape our perception of what is out there for us to achieve, by making visible some things while obscuring others (Ananny and Crawford 2018).

## 3.2 The capability approach as a guide for assessing the ethical dimension of AI tools

The ethical dimension of AI is being redefined as the impact on central capabilities, namely on those substantial freedoms which restriction can violate human dignity. Therefore, an AI tool is ethically controversial (in the soft sense), when it impacts capabilities in a negative way. But what have we gained exactly with this? This formulation seems to suffer from the same applicability problem found within the principlist approach to AI. As it is difficult to assess in practice how AI tools can violate important principles such as autonomy or justice, so it is not clear how certain designs of AI can possibly impact capabilities in a negative way. In other words, it seems that we have replaced a framework with another one suffering from the same flaws, and this time with a much heavier theoretical background.

However, the capability approach comes with the resources to overcome the implementation problem affecting AI principlism (Morley et al 2019; Bezuidenhout and Ratti 2021). In order to understand how AI tools can impact capabilities, we have to consider once again the structure of combined capabilities, i.e. internal capabilities, basic capabilities, conversion factors, and technology. In particular, we see conversion factors as central: an ethical analysis of AI tools based on the capability approach should look at conversion factors in the following way. AI tools are usually advertised or described as being designed to be beneficial in some way to users. In some cases, the benefits can be easily spelled out in terms of the expansion of central capabilities. This happens when AI tools bear directly to the functionings related to important capabilities such as health (e.g. AI tools that promise to predict a prognosis), or control over one's material environment (e.g. when AI tools are used to decide who should get a mortgage) just to name a few. Even when the benefits are not obviously related to a central capability, the context usually can provide indications of which central capabilities might be impacted (e.g. if it is an AI tool used in the context of education, then the central capabilities impacted can potentially be the ones most related to education). However, we know that a technology can expand a capability only when the right conversion factors are in place. If the conversion factors are not there, then the technology will not expand any capability – remember the example of how bicycles can expand capabilities in the Netherlands, but not in a desert. Therefore, the question to ask is the following: *in designing an algorithmic system, which conversion factors do AI practitioners assume that end users must have in order to benefit from the AI?*

To understand the importance of this question, consider a famous case much discussed a few years ago. In (2019), Obermeyer et al. analyze the predictions made by a widely used health-risk algorithm that falsely concluded Black patients were healthier than equally sick White patients, though the race-blind algorithm appeared well calibrated (and thus unbiased) across races. Among the many problems, what emerged was that the algorithm used 'health expenditure' as a proxy for health risk. So, the algorithm appeared fair with respect to health care expenditures but not actually with respect to meeting healthcare needs. It is reasonable to think about 'health expenditure' as a proxy for risk: the more a person spends on health, the more this person is likely to experience health-related problems (with the exception of serious cases of hypochondria). However, it will be a proxy only for those individuals who already have access to healthcare in the first place. 'Health care access' is a conversion factor (Prah Ruger 2010; Venkatapuram 2011) that also depends on other conversion factors, such as having a full-time job; health insurance; a stable income; living in an area where healthcare is accessible; etc. In other words, the algorithm was performing well only for those individuals who have certain conversion factors in place. Therefore, health capability (Prah Ruger 2010) was expanded only for those where the combined capability included specific conversion factors. AI practitioners have assumed that all end-users had a homogeneous level of conversion factors; however, with this controversial assumption, they automatically exclude and make invisible all those who lack that particular level of conversion factors, thereby restricting health capability and, as a consequence, other central capabilities as well (Ratti and Graves 2021). Therefore, to anticipate whether an AI tool will fulfill its promise of capability expansion, one should look at what configuration of conversion factors end-users must have to benefit from the tool.

### 3.3 AI Ethics Through Capabilities

Through the lens of capabilities, AI Ethics becomes the investigation of the impact of AI tools on Nussbaum's central capabilities. Ethical analysis then means asking two specific questions:

1. Which conversion factors should AI practitioners assume that end users must have to benefit from the AI? This question can be answered by means of a philosophical and sociological investigation
2. How are information and assumptions about conversion factors concretely accounted for in the design of AI tools? This requires a technical investigation

The idea is that, given the answers to 1 and 2, one can reasonably anticipate the impact of AI tools on central capabilities. In the next section, we show how this concretely happens in the case of ethics-based auditing.

## 4. ETHICS-BASED AUDITING AND CAPABILITIES

The capability approach, as we have shown in section 3, can be used as a foundational approach to AI ethics. We have also sketched how it can 'operationalize' AI ethics in the sense of soft governance. However, we have not entered much into the concrete details of how the capability approach has a distinctive way of solving the 'from what to how' problem. In this section, we show how through the capability approach we can formulate a well-defined structure for internal ethics-based auditing (EBA). This structure makes AI ethics (as a soft governance tool) concrete (a) by showing to practitioners where the 'ethics' lies in the technical decisions, and (b) by providing concrete suggestions on how to guide the discussions between practitioners and various stakeholders.

### 4.1 Auditing and Ethics-based Auditing

Auditing is generally defined as a systematic process of gathering evidence on the functionality of a certain system, and then communicating the results to a specific audience (Mokander and Floridi 2022). The aim of an audit is to understand whether a system behaves in accordance with the intended (or claimed) functionality.

There are several distinctions that cut across types of audits differently (Mokander et al 2023). For instance, audits can be performed internally by a company, or externally by certified auditors (Raji et al 2020; IIA 2022). There are compliance audits (whose goal is to explicitly compare the behaviors of systems to precise benchmarks), as well as what are called 'risk-based' audits (even though they are not necessarily about risk), which require more open-ended investigations (NIST 2022). Audits can happen at the functional level (i.e. a system purpose), at a technical level (e.g. coding, model development, etc) or at the impact level (e.g. the types and severity of effects of AI deployment to stakeholders). The most comprehensive type of AI audit will cover these three aspects.

Ethics-based auditing (EBA) is an audit process that looks at an AI system's behavior from a specific angle. As the name suggests, EBA investigates the ethical dimension of AI systems (Mokander and Axente 2021). In EBAs, the behavior and the characteristics of AI tools are scrutinized with two goals in mind. First, one objective is to make sure that AI systems align with a set of ethical principles or values (Brundage et al 2020). Second, through EBAs (especially internal) a goal is to stimulate discussions and awareness on a pro-ethical design, rather than just a 'reactive' one (Raji et al 2020). Because of the difficult nature of 'aligning' values and ethical principles to AI systems, EBAs are usually 'risk-based', in the sense that there are no specific benchmarks or measures, and more 'open-ended' questions must be asked. EBAs can be external, but it is through internal audits (which are the ones we focus here) that the second goal can be better achieved.

So far, EBAs have been conceptualized in two different ways. One is based on the AI principlism. The idea is that EBAs should provide evidence that an AI tool aligns with some ethical principles. There are two main problems with this approach, and both jeopardize the first goal of EBAs, which is to make sure that AI systems are in accordance with a set of ethical principles or values. The first shortcoming is that it is not clear how we get past the problem that principles are usually vague and not well-specified, so that a technology company may understand respect for autonomy or explainability in one way, while external auditors in another. But even if we find a solution to this problem, there is still little guidance (the second shortcoming of AI principlism) on what it means to 'identify' principles in algorithmic systems. In fact, it is difficult even in principle to understand what it means for an algorithmic system to 'embed' something like a principle - what does one have to point to in the algorithmic system to argue that there is the principle X?

One solution to these issues - and this is the second dominant approach in EBA - is to measure the impact of AI tools on groups with protected characteristics, in order to establish whether the tool is 'discriminating' (Brown et al 2021; Saleiro et al 2018). This more technical approach is what informs communities like fair-ML (Fazelpour and Danks 2021). By measuring 'fairness', one can satisfy the first goal of EBAs, which is to establish whether an AI tool is in accordance with a principle (in this case, fairness). However, there are two main problems even for this approach. First, it has been shown that there are many measures of fairness, and that some of these are mutually exclusive (see for instance Kleinberg et al 2017). This means that there is no clear way of framing a discussion on which measure of (and hence which definition of) fairness to employ. One can simply use the most convenient off-the-shelf tool depending on the context. But this pragmatic solution often comes with the second problem: uncritically choosing the most convenient 'off-the-shelf' tool jeopardizes the second goal of EBAs, which is to stimulate ethical reflection.

## 4.2 EBAs Through Capabilities

The capability approach can be used to achieve both goals of EBAs, as a result of functional, technical, and impact audits. In a nutshell, EBAs provide an investigation of the ethical dimension of AI tools by establishing how such tools impact central capabilities. In what follows, we focus on the EBAs of medical AI tools.

*4.2.1 Preliminaries of capability-based EBAs*

As shown in Section 3.3, any investigation of the ethical dimension of AI tools understood in capability terms must start from a conceptual investigation comprising both philosophy and sociological components. This also applies to EBAs. In order to answer the first question mentioned in 3.3 for EBAs, two important issues need to be addressed.

First, it is important to understand which capability the AI tool being audited is supposed to expand, or in which capability terms the alleged benefits of a tool can be translated. According to Nussbaum, there should not be any tradeoff between the ten central capabilities. In other words, preserving human dignity means providing an acceptable level of all central capabilities. For the sake of simplifying our analysis, we will just focus on health capability, given the medical context. We are aware that capabilities are interrelated (Oosterlaken 2013), and depending on the specific way the capability of health is impacted, there will be specific downstream effects to other central capabilities (Venkatapuram 2011; Ratti and Graves 2021). However, our purpose here is illustrative, and just focusing on one capability is a good enough proof of concept.

How should we conceptualize health as a capability? Nussbaum calls it 'bodily health' (though the move to have the 'mental' dimension covered by other capabilities is dubious), and it is defined in terms of functionings rather than capability proper. In particular, she defines it as "[b]eing able to have good health, including reproductive health; to be adequately nourished; to have adequate shelter" (2011, p 33). This definition emphasizes the 'accomplishment' dimension of health, and a more substantive characterization as a substantial freedom is missing. A similar characterization comes from Venkatapuram (2011). He calls it 'capability to be healthy', which is seen, at least in part, instrumentally contributing to the other central capabilities, rather than being *per se* a capability. The emphasis is on how "all four components - personal features/needs, behaviours, surrounding social and physical conditions - must be capable of coming together to create practical possibility to achieve each CHC [i.e. central human capability] up to or above the specified thresholds" (2011, pp 154-155), while neglecting the 'freedom' aspect of this central capability. However, there is a sense in which health can be considered a genuine capability. For instance, there have been works in health economics by Jennifer Prah Ruger and her group that have provided a more in-depth characterization of what health capability is. In (2010), she distinguishes between health functioning and health agency. The latter is defined as individuals' ability to achieve health-related goals (or functionings) that they value, in a way that they act as agents of their own health, while the former is the actual health outcome resulting from the actions permitted by health agency. Therefore, we can define health capability as *the opportunity to exercise personal responsibility in health-related matters*.

The second issue to address in the conceptual investigation is how exactly we measure - or at least identify - an impact on health as a capability, or what are the factors that mostly impact something like health agency. The answer lies in the structure of combined capabilities: without certain conversion factors, a capability cannot just be expanded. In this case, we can think about a capability as something that can be broken down into causal components of individual characteristics, needs, and external conditions (Venkatapuram 2013). Individual health capabilities, Prah Ruger may say, depend on how external and internal circumstances add or detract from an individual. In order to investigate the impact on health capability – or, to use Prah Ruger's expression, to create a *health capability profile* - we need to do research about the interaction between individuals and various conversion factors. In particular, the second question requires understanding which configuration of conversion factors end users must have to benefit (i.e. for health capability to be expanded) from the audited AI tool. If you take the case of Obermeyer et al mentioned above, the configuration of conversion factors is the one that allows end-users to have healthcare access in the American context.

*4.2.2 Technical Analysis of AI Systems Through the Lens of Capabilities*

After this philosophical and sociological investigation, a more technical investigation is necessary. As soon the configuration of conversion factors needed by end-users to have their health capability expanded is established, one has to identify how a particular AI system, in fact, interacts with data about conversion factors. 'Interacting' here means how, for instance, an AI tool cuts out certain individuals on the basis of thresholds based on data about the conversion factors themselves; how the decisions of those who design the system include or exclude important information about conversion factors, and so on. This is a 'technical' investigation, where 'technical' here means that auditors should investigate how the technical aspects of algorithmic systems and how they have been built impact capabilities, in particular health capability.

Consider a concrete example. Imagine a case of a company that is trying to improve health outcomes by delivering customized nutritional meals to patients with diabetes. The company would work with payors and healthcare providers to define criteria to be used for patients to receive meal delivery. An AI tool is then developed to process electronic health records (EHR) data in order to select those patients that may better benefit from the program. After preliminary development, the tool is to be audited internally - audited for capabilities. As we have argued above, in order to do this, one has to decompose capabilities into their constituent parts, and then identify especially the kind of conversion factors required for a capability to be expanded by the AI tool under audit. In other words, to understand the impact on capabilities, one has to analyze conversion factors, and how they interact with AI tools. But how do we identify 'conversion factors' in AI tools *exactly*? Conversion factors are 'things' out there in the world, and not in algorithmic systems. Let us see how.

In our diabetes example, it is possible to imagine a typical data science pipeline (see Figure 1), that practitioners will have to go through to build and use the tool. Conversion factors are then 'addressed' by practitioners both in the form of data they use to train the algorithm, as well as in the implicit assumptions about them in executing various technical tasks. We can briefly go through some of them.

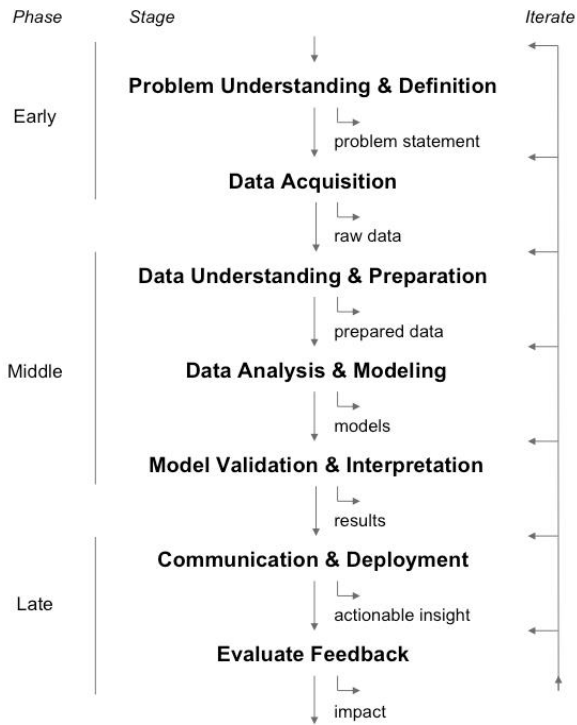

Figure 1. A typical data science pipeline (adapted from Ratti and Graves 2021)

Let us start with the stage of problem understanding and definition. This is where capabilities can most clearly be tackled directly. In this stage, practitioners discuss the nature and the goals of the system. The goal of an AI tool used to select the right patients of a nutritional program will likely be discussed in terms of health outcomes, but health agency is important as well. Because adults generally make their own choices about what they eat, just delivering nutritious meals without considering health agency is ineffective. Inclusion and exclusion criteria would incorporate clinical measures, such as blood sugar levels, but can also include proxies for motivation and other psychological and social factors relevant for health agency. Being explicit about conversion factors is beneficial. Those who lack transportation to places with nutritious food would be plausible candidates, while assuming everyone has freezer space and a microwave oven may be invalid for those homeless or with insecure housing.

During the stage of data acquisition, data relevant to conversion factors are likely to be central. For instance, what features of data are useful to understand the personal, social, and environmental factors of data subjects? How are these features considered by practitioners? Another example of an important aspect of data acquired is their quality. In fact, data quality will tell us a lot in terms of its provenance, and about the conversion factors of those individuals whose data are collected. For instance, in selecting unstructured EHR data, there is a tendency (motivated by practical convenience) to use data that is exported cleanly from electronic medical records at the expense of data requiring intense optical character recognition (OCR). This is because OCR data are full of gaps and mistakes, and they require much more labor to be processed and used. This is a strategy that is often endorsed explicitly. For instance, in building MEDomics (i.e., a learning infrastructure through which health data are organized with the goal of applying AI tools for various health-related goals) Morin et al (2021) explicitly say "we selected individuals with high-quality data, defined as high-volume, high-continuity and highly complete feature sets (...) for further development of statistical model" (p 711). But it is worth considering that data in OCR may come from underserved area, and hence by not using OCR data one might exclude those individuals coming from those underserved areas, with the risk of excluding data subjects with a specific combination of conversion factors, which will be different from the combination of conversion factors of those coming from the 'high-quality' data areas. Attention should be paid to the conversion factors shaping health agency and that can be found in or inferred from data sets that one chooses to use.

Similar issues emerge in the other stages. In data understanding and preparation, the way missing data are handled can make a difference in terms of capability impact, as shown in (Ratti and Graves 2021). In fact, considering data as missing at random or not at random should be informed by an understanding of conversion factors. If a subject has missing or irregularly reported results, it matters whether that occurs due to lack of transportation to a clinic, poor adherence to medical advice, or careful but unreported self-monitoring using personal medical devices. In the stage of data analysis and modeling, attention should be paid to how conversion factors are hidden or made salient by the modeling strategy, from simple descriptive statistics, to machine learning techniques - for instance, in addition to ensuring representative samples, one should consider if adequate features are used, with Obermeyer's (2019) recognition that healthcare cost is an inadequate proxy for healthcare risk or need a pertinent example. More subtly, if one uses dimensionality reduction, then attention should be paid to how 'collapsing' some data dimensions does not create a loss in information about conversion factors, and hence capabilities. Healthcare access, which affects numerous clinical criteria, depends upon many factors, and unless explicitly modeled may be lost as noise among more simply defined factors. In model validation and interpretation, when comparing outputs to independent observations, one will judge whether the model is performing as expected. But, as shown in (Graves and Ratti 2022), there is no simple and straightforward measurement; in fact, one has to tune the model to maximize some measurement or balance some tradeoff. One might use follow-up surveys or participant engagement to evaluate the meal program's contribution to any clinical improvements, but one still has to determine and tradeoff this intervention's benefit to participants compared to alternative options, such as merely providing nutritional information, and the program cost for those who did not benefit plus the additional healthcare cost for those who would have benefited but were not selected. Identifying thresholds for care may beneficially help constrain the overall optimization problem to a more tractable one, such as ensuring that everyone has access to one nutritious meal per day.

These are just a few examples of how conversion factors can, metaphorically, interact with AI tools.

*4.2.3 A Structure for Internal EBAs*

From what we have said in 4.2.1 and 4.2.2, one can use the conceptual and technical investigations in tandem to identify the impact of AI tools on capabilities. In order to make this more concrete, we envision a four-step procedure for internal EBAs.

In the first step, after the initial preliminary development of an AI tool, AI practitioners are asked to compile a detailed document on how they have built the tool. This should follow the pipeline we have sketched in 4.2.2 and Figure 1, and include also other standardized documents like data sheets or model cards (Gebru et al 2018; Mitchell et al 2019).

In the second step, this documentation is analyzed. Because such documents can be long, detailed, and plenty of details may not be relevant to capabilities, we envision the development of a software that can be used to flag in the documentation relevant *loci* where the way the system has been built deal with data about conversion factors or presuppositions about conversion factors. Information retrieval and natural language processing (NLP) techniques can identify locations in the document. For this, the internal auditor defines the terms and other descriptors relevant for the capability audit, in collaboration with other experts. The tool identifies locations in the main and supplementary documents where those terms and descriptions occur. Using an AI tool in the audit of other AI tools supports efficient auditing of large, complex development projects, and enables deriving quantitative measures for comparison and long-term tracking.

In the third step, the results are presented to relevant internal auditors. There is a trend in recent literature to promote interdisciplinary and participatory design of AI tools (Ratti and Graves 2021). Our EBAs structure is interdisciplinary in nature, as internal auditors should be either ethicists or social scientists (or both). Internal auditors will analyze the results provided in the second phase, and they will integrate these with the conceptual investigation about conversion factors required for the AI tool to expand capabilities. In this phase, tools for fair-ML can be used also to get an understanding of how specific groups are potentially excluded by the tool, such as racial or gender bias that may indicate social or economic factors affecting capabilities.

Finally, internal auditors discuss the results of their analysis with practitioners, with the aim of modifying the design of the algorithmic system to make it more beneficial to capabilities. The discussion can take the form of a workshop, where people can brainstorm about the results. For instance, one may discuss how the goal of the system has been formulated with respect to capabilities. Having access to nutritional food is indeed something that can possibly expand health agency. But if the successful intervention is measured in terms of the self-glucose test that people with diabetes would have to take, then maybe we are assuming a certain level of health self-knowledge (which is a conversion factor) that few people have. Or in the case of OCR and EHR data, auditors and data scientists may discuss if there are differences in EHR and OCR with respect to environmental conversion factors, such as facilities, and resources of neighborhoods, social security of the macrosocial environment, presence of barriers to health facilities, and general effectiveness of those facilities. In the case of missing data, one can discuss if it is a matter of healthcare access whether some data are missing and consider the factors that would affect such access. This is the most participatory and interdisciplinary part of the audits, and it can foster an inclusive design of algorithmic systems that can bring benefits to capabilities expansion.

## 5. CONCLUSION

In this article, we have proposed to see the ethics of AI through the lens of capabilities. We have shown that re-conceptualizing AI ethics through the capability approach disambiguates theoretical concerns regarding AI ethics as a discipline, as well as addressing practical concerns around existing frameworks. We have made our case by showing how this idea can guide EBAs for medical AI tools.

To conclude, we think that a capability-based AI ethics need not necessarily be in contrast to principlist and more technical solutions (e.g. the ones provided by fair-ML community). We can just say that the principlist approach is one piece of the puzzle of hard regulation, and it does not need to be employed everywhere. When it comes to soft governance of AI practice, an approach that is more attentive to the idiosyncratic and micro-incremental choices of the daily activities of AI practitioners is more suitable. We have shown that our capability-based AI ethics is a valid candidate. Our approach can also be used in tandem with more technical approaches, as it can provide the resources for the ethical reflection that is needed to choose, for instance, among different fair-ML tools in the first place.

**NOTES**

**Acknowledgements**: authors would like to thank audiences at conferences, seminars, and workshops where this work has been presented. In particular, audiences at the University of Bielefeld ZiF (workshop "The Epistemology of Evidence-based Policy"); audiences at an IDSAI seminar in Exeter; audiences at the Notre Dame-IBM Technology Ethics Lab. This work has been supported by Innovate UK (grant n. 10063119), HORIZON EUROPE Framework Programme (grant n. 10063119), and University of Notre Dame-IBM grants (n. 262812UB and n. )

[1] For an overview of the limits of this strategy, see Mittelstadt (2019).
[2] In all articles taking stock of the capability approach that we have consulted, the distinction plays a central role (Oosterlaken 2012; Robeyns 2005; Alkyre 2005)
[3] Nussbaum still thinks that it is important to distinguish between combined capabilities and internal/basic capabilities, because the distinction also reflects a distinction between two (partially overlapping) tasks of governments.
[4] While revising this article, we noted that scholars have started to explore the synergies between the capability approach and AI ethics (Buccella 2023; London and Heidari 2023). These works have different goals than the ones that we have in the present manuscript. For instance, Buccella's goal is to emphasize the importance of social justice in the deployment of AI, while London and Heidari look at CA as an instrument to formalize ethical concerns.